\documentclass[epj,final]{svjour}
\usepackage{graphics}
\begin{document}
\title{\boldmath
  Probing hypernuclei at $\overline{\mbox{P}}$ANDA and at 
  MAMI-C\thanks{Supported by the Bundesministerium f{\"u}r
  Bildung und Forschung (bmb+f) under contract no.\ 06MZ176 and
  by the EU integrated infrastructure initiative
  HadronPhysics project under contract no.\ RII3-CT-2004-506078.}}

\author{P.\ Achenbach}
\institute{Institut f\"ur Kernphysik, Johannes Gutenberg-Universit\"at,
  D-55099 Mainz, Germany.}
\date{Societ\`a Italiana di Fisica $/$ Springer-Verlag 2007}
\abstract{ 
  Spectroscopy of $\Lambda$ hypernuclei has recently become one of the
  most valuable tools for the experimental investigation of
  strangeness nuclear physics. Several new approached are being
  pursued currently: In Mainz, the Microtron MAMI has been upgraded to
  1.5\,GeV electron beam energy and will be used to produce strange
  hadronic systems in the near future.  The KaoS spectrometer is being
  installed for large acceptance, high resolution strangeness reaction
  spectroscopy at the existing spectrometer facility.\\
  The Mainz hypernuclei research programme will be complemented by
  experiments on multi-strange systems at the planned FAIR facility at
  GSI. The $\gamma$-ray spectroscopy of double $\Lambda$ hypernuclei
  produced via $\Xi\bar{\Xi}$ pair production is one of the four main
  topics which will be addressed by the $\overline{\mbox{P}}${ANDA}\
  Collaboration. In this paper the status of the planned experiments
  and the future prospects are presented.
} 
\maketitle
%


\section{Introduction}
\label{sec:intro}
Open strangeness production is undergoing a renewed interest, both
theoretically and experimentally. On the theoretical side, due to the
non-perturbative nature of Quantum Chromodynamics (QCD) at low
energies, strange systems cannot be described by the fundamental
equations for the dynamics of (asymptotically free) quarks and gluons.
Instead, isobaric models are commonly used in kaon production, where
the hadrons are treated as effective degrees of freedom. The partonic
constituents can also be considered along the lines of chiral models,
which take the chiral symmetry of the QCD Lagrangian and its
spontaneous breakdown into account. Although lattice QCD calculations
are not yet relevant to kaon production, it is anticipated that
precise experimental data on strangeness production will challenge and
improve our understanding of the strong interaction in the low energy
regime of QCD. Since direct scattering experiments with hyperons are
impractical, the spectroscopy of hypernuclei provides a unique
approach to explore the baryon-baryon interaction. This field of
physics will be addressed by the Mainz Microtron MAMI, recently
upgraded to 1.5\,GeV electron beam energy and at the antiproton
storage ring HESR of the future FAIR facility at GSI.

The $(K^{-},\pi^{-})$ reaction is characterised by the existence of a
"magic momentum" where the recoil momentum of the hyperon essentially
becomes zero. It populates, consequently, substitutional states in
which a nucleon is converted to a hyperon in the same orbital state.
The $(e,e',K^{+})$ reaction, in contrast, produces hypernuclei by
converting a proton into a hyperon and transfers a large momentum to a
hypernucleus. The electroproduction process has the unique
characteristic of providing large amplitudes for the population of
spin-flip hypernuclear states with unnatural
parities~\cite{Motoba1994}.

In the case of the $\Lambda N$ strong potential, the spin-orbit term
is known to be much smaller than in $NN$ interaction.  Although these
transitions can only be observed using
$\gamma$-spectroscopy~\cite{Tamura2005}, reaction spectra are equally
important because they provide the complete spectrum of excitations
with a strength that not only gives the spectroscopic factor, but the
transition matrix. A deeper understanding of the $\Lambda$ potential
relies on obtaining high quality data for hypernuclei. The present
experimental data on binding energies and detailed spectroscopic
features are limited in quantity and quality and refer mostly to light
($s$- and $p$-shell) hypernuclei.


\section{Production of hypernuclei at MAMI-C}
\label{sec:MAMI-C}
Electron beams have excellent spatial and energy definitions, and
targets can be physically small and thin ($10-50$\, mg$/$cm$^2$)
allowing studies of almost any isotope.  The small cross section for
the reaction, $\sigma\sim 140$\,nb$/$sr on a $^{12}${C}
target~\cite{Miyoshi2003}, compared to strangeness exchange
n$(K^{-},\pi^{-})\Lambda$ or to associated production
n$(\pi^{+},K^{+})\Lambda$ is well compensated by the available high
electron beam intensities. Even though an increasing number of new
experiments are now being performed in hypernuclear spectroscopy, our
knowledge on hypernuclei is still limited to a small number of
isotopes. With the commissioning of the 1.5\,GeV electron beam of
MAMI-C the study of hypernuclei has become possible in Mainz.

In electroproduction the angular distribution of kaons associated with
a given hypernuclear state is sensitive to the $\Lambda$ wave function
inside the nucleus~\cite{Shinmura1994,Bennhold1999}. A hypernuclear
$\Lambda$ samples the nuclear core where there is little direct
information on the single particle structure. In addition, high
resolution spectroscopic studies of hypernuclei can be performed in
order to provide the most valuable experimental information on the
$\Lambda$ dynamics in nuclei.

One of the factors decisive for the feasibility of an experiment at
MAMI is of course the reaction rate. To optimise the reaction rate the
hypernuclear formation in impulse approximation was modelled. In order
to form a hypernucleus, the hyperon produced in the reaction has to be
bound by the core nucleus. The transition form factor depends very
much on the transferred momentum to the hyperon. If the momentum
transfer is large compared with typical nuclear Fermi momenta, the
hyperon will emerge from the nucleus.

Assuming 3-momentum conservation at the vertices of the Feynman
diagram in impulse approximation the transferred momentum can be
written as the difference between $\Lambda$ momentum,
$\vec{p}_{\Lambda}$, and core nucleus momentum, $\vec{p}_{A-1}$, and
is a function of the momentum of the virtual proton, $\vec{k}$, and
the recoil momentum of the hypernucleus, $\vec{p}_Y$: $q(k) \equiv
\big|\vec{p}_{\Lambda}-\vec{p}_{A-1}\big| = \big|\vec{p}_Y +
2\vec{k}\big|$. For the kinematical optimisation an approximate Fermi
Gas distribution for the virtual proton, $F=
2\pi\int_0^\infty\!n(k)k^2\,d k$ was assumed, where the distribution
function, $n(k)$, is Gaussian, $n(k)= (2^{-4}k_F\sqrt{\pi})^{-3}
\exp{-\sqrt{2}k^2/k_F^2}$. By modelling the transition form factor
with an exponential function, $S_\Lambda=
\int\!\!\!\int\exp(-q(k)/\sigma_p) n(k)\, k^2\,d k\,d\cos\theta$, it
can be evaluated as a function of the kinematic variables. Since
electroproduction is a high momentum transfer reaction similar to
$(\pi^+,K^+)$, the transition form factor takes a minimum at threshold
and increases as the virtual photon energy increases.  An
interpretation of the transition form factor in terms of a sticking
probability is given, for example, in~\cite{Bando1990}, for the case
of harmonic oscillator wave functions. Using this interpretation the
sticking probability prevails at low $q$ and substitutional states are
exclusively populated, whereas at $q\sim k_F$, high quantum numbers
and spin stretched states are favoured.

\begin{figure}[htb]
  \resizebox{\columnwidth}{!}{
    \includegraphics{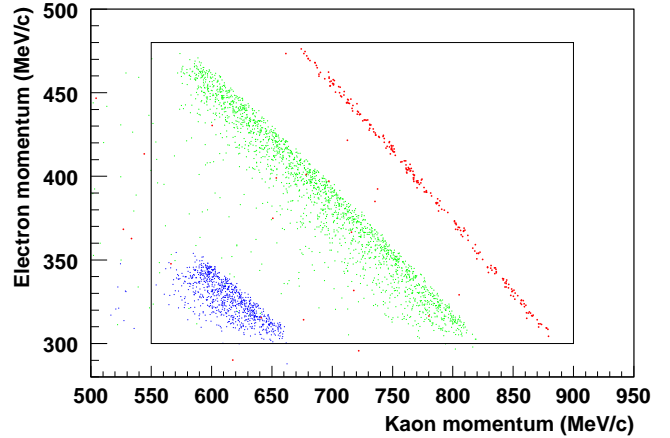}}
  \caption{Simulated correlation between electron and kaon momenta,
    where $\Lambda$ (green) and $\Sigma$ (blue) hyperons have been
    generated for the elementary production off the proton and the
    $^{12}_{\Lambda}$B hypernuclei (red) have been generated in a
    carbon target. The rectangular box indicates the acceptance of
    KaoS.}
  \label{fig:correlation}
\end{figure}

The cross section for electroproduction can be written in a very
intuitive form by separating out a factor, $\Gamma$, which multiplies
the off-shell (virtual) photoproduction cross sections. This factor
may be interpreted as the flux of the virtual photon field per
scattered electron into $d E' d\Omega$ and can be written as $\Gamma =
{\alpha \over 2 \pi^2} {E' \over E} {k_\gamma \over Q^2} {1 \over 1 -
  \epsilon}$, with $k_\gamma= (W^2-m_i^2)/2m_i$ and
$\epsilon=\big(1+2\frac{{\vert \mathbf{q}\vert}^2}{Q^2}{\tan^2
  \frac{{\theta}}{2}}\big)^{-1}$, where $\mathbf{q}$ is the virtual
photon three-momentum.  The virtual photon flux factor has the feature
that it is very forward peaked. While the flux factor for a fixed
photon energy, $\omega= E_e-E_e'$, increases with beam energy, so does
the mass resolution and the background rate. In
Fig.~\ref{fig:correlation} the electron momentum is plotted versus the
kaon momentum, where $\Lambda$ and $\Sigma$ hyperons have been
generated for the elementary production off the proton and the
$^{12}_{\Lambda}$B hypernuclei have been generated in a carbon target.
The events have been generated randomly in phase-space and weighted by
a factor for the virtual photon flux and the modelled transition form
factor. In the Monte Carlo, the production probability was assumed to
drop exponentially with the relative momentum between $\Lambda$ and
core nucleus and typical values of $\sigma_p=$ 100\,MeV$/c$ and $k_F=$
200\,MeV$/c$ were assumed. The proposed kinematic setting for a first
hypernuclear formation experiment has been defined accordingly. From
the acceptance at a magnetic field setting of 1.1\,T a scattered
electron energy of 400\,MeV at a scattering angle of 3$^\circ$ was
chosen, resulting in 1100\,MeV photon energy. This correlates to a
kaon momentum of about 760\,MeV$/c$.


\subsection{The KaoS spectrometer}
\label{subsec:kaos}
Experimentally, the conservation of strangeness in electromagnetic and
strong interactions allows the tagging of baryonic systems with open
strangeness, e.g.\ baryon resonances or hypernuclei, by detecting a
kaon in the final channel.

\begin{figure}
  \resizebox{\columnwidth}{!}{
    \includegraphics{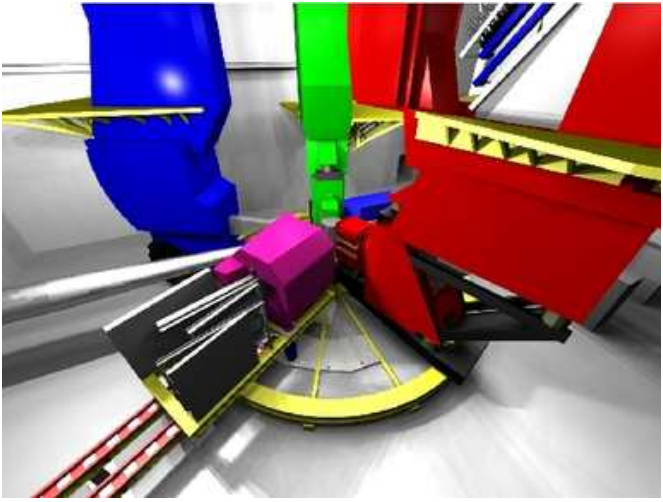}}
  \caption{Graphical visualisation of KaoS at its measurement position
    close to the target in the spectrometer hall.}
  \label{fig:kaos}
\end{figure}

KaoS is a very compact magnetic spectrometer suitable especially for
the detection of kaons. It was built for the GSI for heavy ion induced
experiments~\cite{Senger1993}. During May and June 2003 the
spectrometer was brought to Mainz. A graphical visualisation of KaoS
at its future measurement position close to the target in the
spectrometer hall is given in Fig.~\ref{fig:kaos}. As a pilot
experiment on kaon electroproduction, the separation of transverse and
longitudinal structure functions in parallel kinematics is planned
with the available detector package~\cite{PAC2003}. The use of KaoS as
a two arm spectrometer for the electroproduction of hypernuclei at
MAMI requires the detection of the scattered electron at laboratory
angles close to 0$^{\circ}$. The kaon detector in the focal plane has
to cover a wide range of scattering angles around 5$^{\circ}$ in
coincidence to the electron detection. In order to cope with the
kinematical situation in electroproduction of hypernuclei and with the
high background rates, KaoS will be equipped with new read-out
electronics, a completely new focal plane detector package for the
electron arm, and a new trigger system.


\subsection{Development of Detectors}
\label{subsec:det}
\begin{figure}[htb]
  \resizebox{\columnwidth}{!}{
    \includegraphics{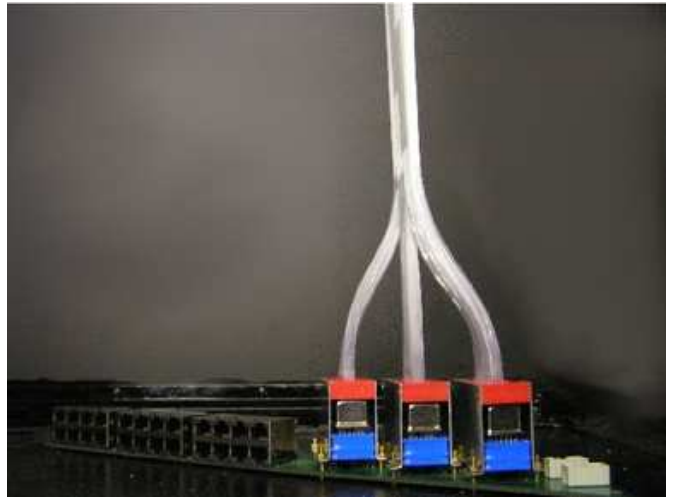}}
  \caption{Photograph of a detector with three joined fibre bundles
    and three MaPMTs mounted on a single front-end board.}
  \label{fig:triple-detector}
\end{figure}

The main focal plane detector of the KaoS electron arm will mainly
consist of 2 horizontal planes of fibre arrays, covering an active
area of $1780 \times 300$\,mm$^2$, and comprising $\sim$2000 channels
per plane. Each plane is divided into detector segments which consist
of 384 fibres in three joined fibre bundles coupled to three
multi-anode photomultipliers (MaPMTs) mounted on a single 96-channel
front-end board, see Fig.~\ref{fig:triple-detector} for a photograph.

A fibre doublet structure is formed from two single layers of fibres,
with one of the fibre layers off-set relative to the other by half a
fibre spacing. The virtue of this configuration is the high fraction
of overlapping fibres, i.e.\ a high detection efficiency, and a small
pitch leading to a good spatial resolution. Two of such double layers
are placed together in the focal plane detector to increase its
efficiency.

The fibres are of type {\sf Kuraray} SCSF-78 with a double cladding of
$0.83$\,mm outer diameter. The total cladding thickness is $0.1\,$mm,
leading to a 0.73\,mm core of refractive index $n_{\it core}=$ 1.6.
Four fibres are grouped to one channel and brought to one pixel of the
photo-tubes of type R7259K from {\sf Hamamatsu} with a 32-channel
linear array of anodes. The photocathode material is bialkali and the
window is made of 1.5\,mm thick borosilicate glass. The effective area
per channel is $0.8 \times 7$\,mm$^2$ with a pitch of 1\,mm.

Instead of supplying dynode voltages through a voltage divider, the
MaPMTs are powered by individual Cock\-croft-Wal\-ton bases,
manufactured by {\sf HVSys}, Dubna. The dc voltage is pulsed and
converted with a voltage doubler ladder network of capacitors and
diodes to higher voltages. The principal advantage is that there is no
need for stiff high voltage cables, since only $\sim 140$\,V has to be
delivered to the first front-end board, where the voltage is
daisy-chained to the other boards of the detector plane. One drawback
is that the inter-dynode voltages can only be equally spaced, which is
acceptable for their actual use.

Since the scattered electrons have an inclination angle of
$50-70^\circ$ with respect to the normal of the focal plane a
configuration with hexagonal fibre packing has been chosen. This
packing, in which each fibre is surrounded by 6 other fibres, has the
highest possible density of $\frac{\pi}{\sqrt{12}}\simeq 0.9$. The
main consequence of the incident angle distribution is an increased
channel multiplicity. The amount of scattering depends primarily on
the momentum and the effective thickness of the scattering medium
(radiation length $X_0= 42\,$cm for polystyrene). Electron
trajectories for different momenta cross the focal plane detector at
different positions under slightly different angles. The average
thickness and its variation was simulated for to be $x= (4.7 \pm
1.3)$\,mm. This number translates into an rms width of $\theta_0=
0.23^\circ$ for $p= 300$\,MeV$/c$. The main consequence of the
incident angle distribution is an increased channel multiplicity.


\subsection{Development of Electronics}
\label{subsec:elect}
\begin{figure}[htb]
  \resizebox{\columnwidth}{!}{
    \includegraphics{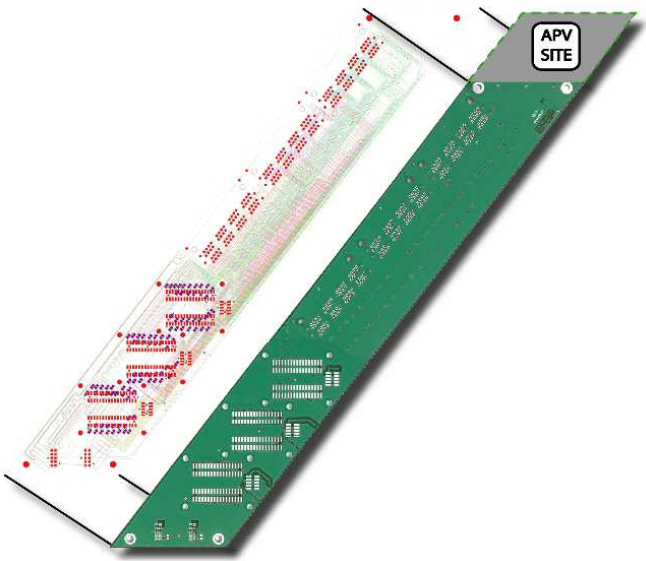}}
  \caption{Photograph and circuit scheme of the triple front-end board
    showing the three MaPMT sockets at the lower left, and the output
    sockets. The site for the APV chip connector is indicated.}
  \label{fig:fe-board}
\end{figure}

\begin{figure}[htb]
  \resizebox{.5\columnwidth}{!}{
    \includegraphics{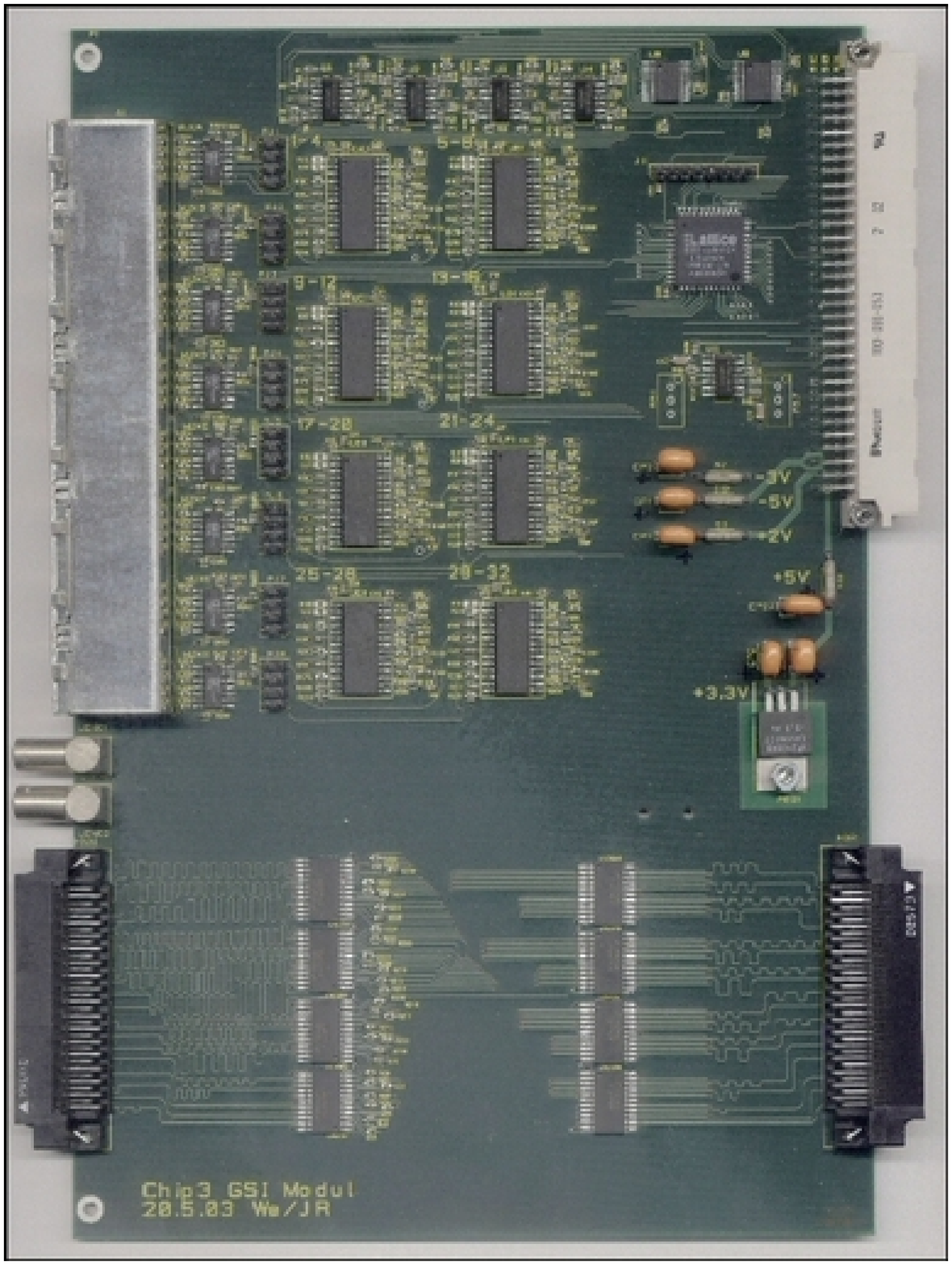}}
  \resizebox{.475\columnwidth}{!}{
    \includegraphics{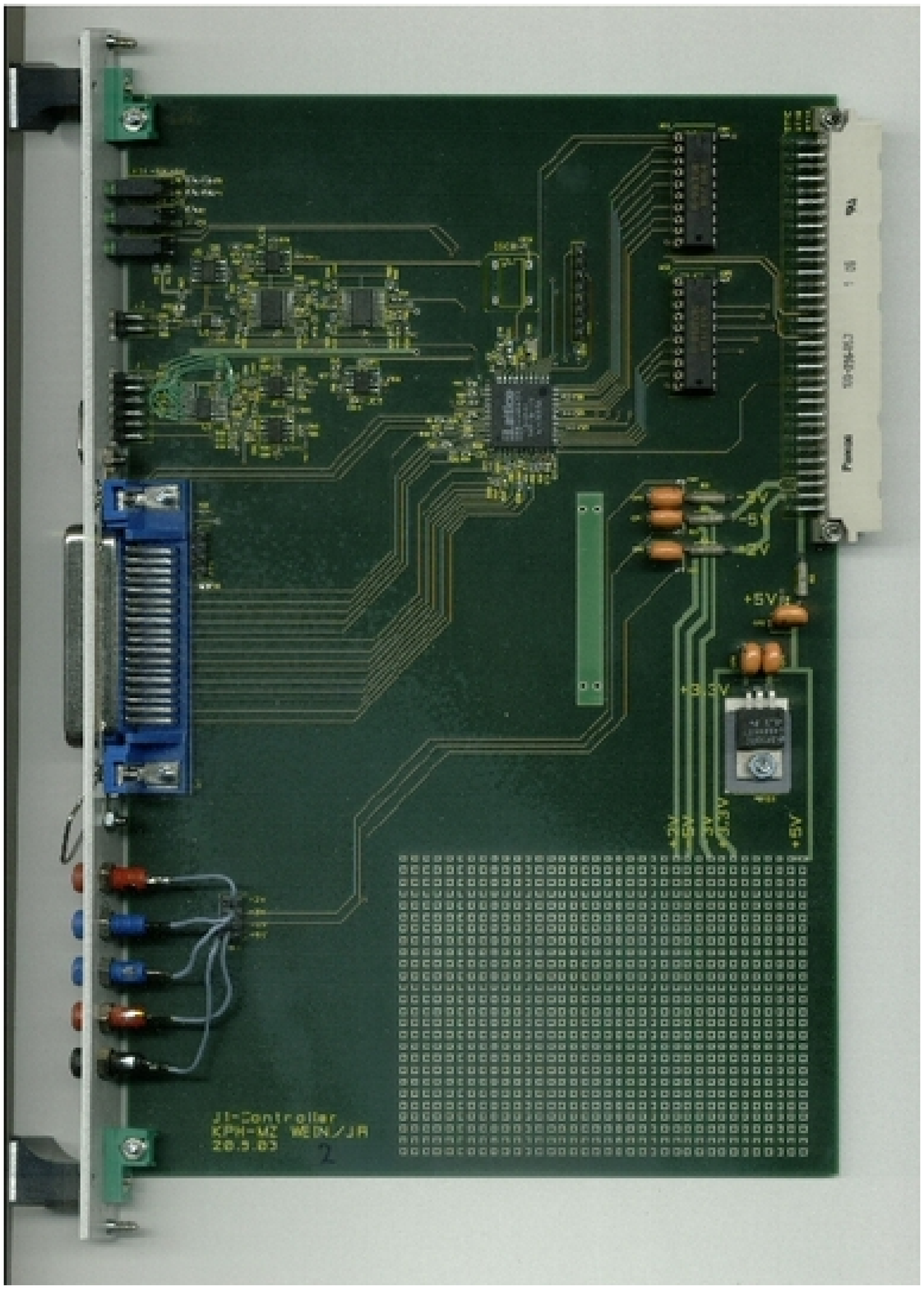}}
  \caption{Left: photograph of the double threshold discriminator
    board.  The two LVDS outputs are visible near the bottom of the
    module. The RJ-45 inputs are located at the top left of the module
    and the two coaxial 50\,$\Omega$ outputs at centre left. Right:
    photograph of the controller board for up to 20 double threshold
    discriminator boards.}
  \label{fig:frontends}
\end{figure}

A 32-channel discriminator board with 4 integrated low-walk double
threshold discriminators (DTDs) for amplitude compensated timing was
built, see Fig.~\ref{fig:frontends}\,(left) for a photograph of the
DTD board. The detectors are connected to the board by RJ-45 cables
(with 4 channels per cable). The DTD boards have two multiplexed
coaxial 50\,$\Omega$ analogue outputs for debugging and two LVDS
outputs, one to be connected to TDC modules, and one to the trigger
modules. A 32-channel analogue output board can be attached to the
discriminator board for a complete analysis of analogue signals. Up to
20 DTD boards fit into a VME 6U crate together with a controller
board, see Fig.~\ref{fig:frontends}\,(right) for a photograph.

Time measurements are performed with an electronic system based on
CATCH, the COMPASS Accumulate, Transfer and Control Hardware system
from the COMPASS experiment at CERN~\cite{COMPASS}. Each CATCH module
is equipped with 4 TDC mezzanine cards with 32 read-out channels per
card. The CATCH module not only serves as an interface between
front-end electronics and the data taking processor, but also acts as
an integral part of the trigger distribution and time synchronisation
system (TCS). At the heart of a mezzanine card is the so-called ${\cal
  F}$1 chip with 8 channels of $\sim 120$\,ps resolution (LSB) each.
The tracking trigger will be derived with VME Universal Processing
Modules. Such a module is equipped with one 500\,MHz FPGA, one 1\,GHz
DSP, and 256 LVDS inputs or outputs.

In addition to timing and trigger, an analogue signal read-out system
capable of handling high channel counts is also needed.  For this
purpose, the APV chip and the GeSiCa data collector card of the
COMPASS electronics is under investigation. The APV chip
acts as an analogue ring-buffer at 40\,MHz, which on demand multiplexes the
appropriate 128-channel sample and sends it to an attached ADC. The
GeSiCa module provides the TCS information processing, data
collection, data concentration and data transfer to PC-based read-out
buffer cards via SLink.

%


\subsection{Performance of Detectors and Electronics}
\label{subsec:GSI}
After a first characterisation of several detector
prototypes~\cite{Achenbach-SNIC06}, a triple detector with 96 read-out
channels for three fibre bundles was tested in Cave~C of GSI with a
$^{12}$C beam of 2\,$A$\/GeV energy provided by SIS18. In each bundle
128 fibres were packed in 4 double layers with a pitch of 0.6\,mm. Two
bundles were aligned to a single plane perpendicular to the beam, and
one bundle formed a parallel plane directly behind.

\begin{figure}
  \resizebox{.75\columnwidth}{!}{
    \includegraphics{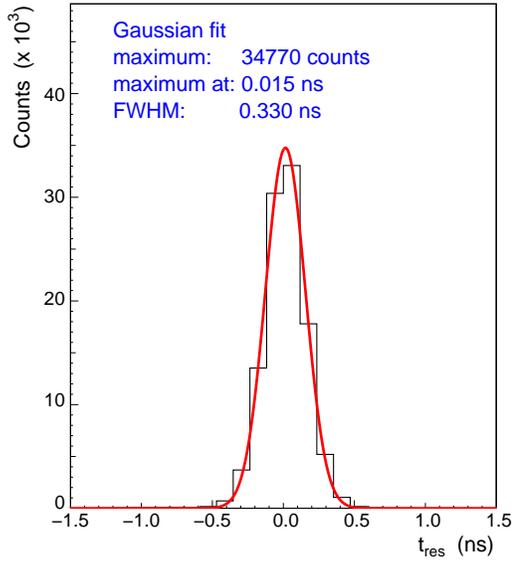}}
  \caption{The residual of hit times between both detector planes.  A
    Gaussian fit is shown with a FWHM of 330\,ps.}
  \label{fig:timeresidual}
\end{figure}

In deducing the time resolution clusters of correlated hit times were
searched for. The cluster with the time closest to the trigger signal
time was retained, and within the cluster the time of the first
arrived signal was chosen as hit time. The hit time residual, defined
as the difference between the two hit times in the two planes of
fibres, was distributed with a FWHM of 330\,ps as shown in
Fig.~\ref{fig:timeresidual}. The time resolution of a single detector
plane was determined to be of the order of FWHM $\sim$
330\,ps$/\sqrt{2}=$ 230\,ps.

\begin{figure}
  \resizebox{.75\columnwidth}{!}{
    \includegraphics{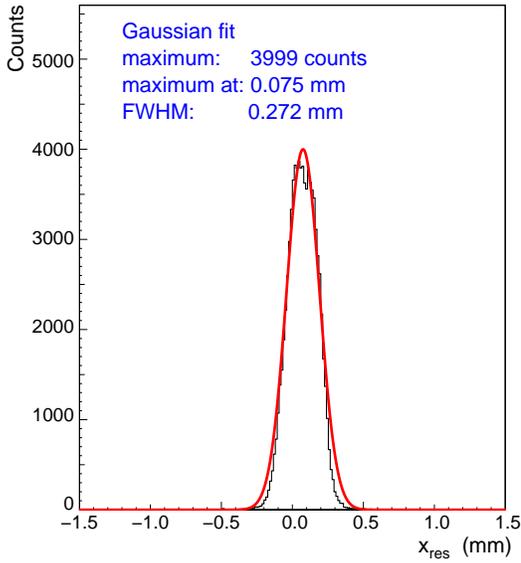}}
  \caption{The residual of beam position estimates between both
    detector planes, using the centroid of charges.  A Gaussian fit is
    shown with a FWHM of 0.27\,mm.}
  \label{fig:spaceresidual}
\end{figure}

Cross-talk in the MaPMT entrance window perturbed the reconstruction
of the particle position. By using the pulse height information the
hit channel was estimated by the centroid of charges of up to 5
neighbouring channels. The hit position residual, defined as the
difference between the two estimates in the two planes of fibres, had
a FWHM of 0.27\,mm as shown in Fig.~\ref{fig:spaceresidual}, although
the estimator contained a spatial digitalisation error. The achieved
accuracy was sufficient for an unambiguous identification of the hit
channel leading to a spatial accuracy of 0.6\,mm$/\sqrt{12}\approx$
170\,$\mu$m (rms).

\subsection{Perspective}
\label{subsec:persp}
The special kinematics for electroproduction of hypernuclei requires
the detection of both, the associated kaon and the scattered electron,
at forward laboratory angles. While the scattered electrons have to be
detected at very forward angles, the kaon detector has to cover a
broader range of up to 15$^{\circ}$ in order to extract dynamical
information from the $K^+$ angular distribution. The potential of
electroproduction at MAMI-C is an energy resolution of a few hundred
keV with reasonable counting rates up to at least medium weight
hypernuclei. An important goal for measuring the excitation spectra
and decay properties of strange hypernuclei is to test the energies
and wave functions from microscopic structure models and to put
constraints on baryon-baryon interaction models.


\section{\boldmath Hypernuclear $\gamma$-spectroscopy with
  $\overline{\mbox{P}}$ANDA}
\label{sec:panda}
At FAIR a new technique for producing double hypernuclei with
antiprotons will be pursued. Their study will reveal the
$\Lambda\Lambda$ strong interaction strength, not feasible with direct
scattering experiments. $\overline{\mbox{P}}$ANDA will also contribute
to the intensive search for the H dibaryon composed of two u, d, and s
quarks, producing a deeply bound system.

At the antiproton beam relatively low momentum $\Xi^-$ can be produced
in $\mathrm{\overline{p}p} \rightarrow \Xi^- \overline{\Xi}^+$ or
$\mathrm{\overline{p}n} \rightarrow \Xi^- \overline{\Xi}^0$
reactions~\cite{PANDA}. The advantages as compared to the kaon induced
reactions are (i) the retainment of the antiproton beam in a storage
ring, (ii) the presence of the $\overline{\Xi}$ antiparticle, that
will be used to tag the strangeness in the reaction, and (iii) the
high production rate with respect to the annihilation of antiprotons
at rest with subsequent strangeness exchange. This allows a rather
high luminosity even with very thin primary targets.

The associated $\overline{\Xi}$ will undergo scattering or (in most
cases) annihilation inside the residual nucleus and a trigger can be
based on the detection of the surviving anti-hyperons under small
angles. Since strangeness is conserved in the strong interaction and
the annihilation products contain at least two anti-kaons, an
alternative tag for the reaction can be based on the detection of
positive kaons with relatively low momentum in
$\overline{\mbox{P}}${ANDA}.

\begin{figure}
  \resizebox{\columnwidth}{!}{
    \includegraphics{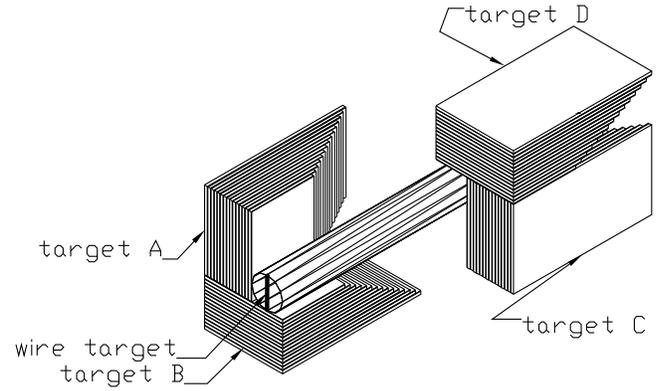}}
  \caption{Exploded view of the target region showing the wire target,
    the beam-line, and the sandwich structure of silicon layers with a
    combination of target materials (A--D).}
  \label{fig:target}
\end{figure}

\begin{figure}
  \resizebox{\columnwidth}{!}{
    \includegraphics{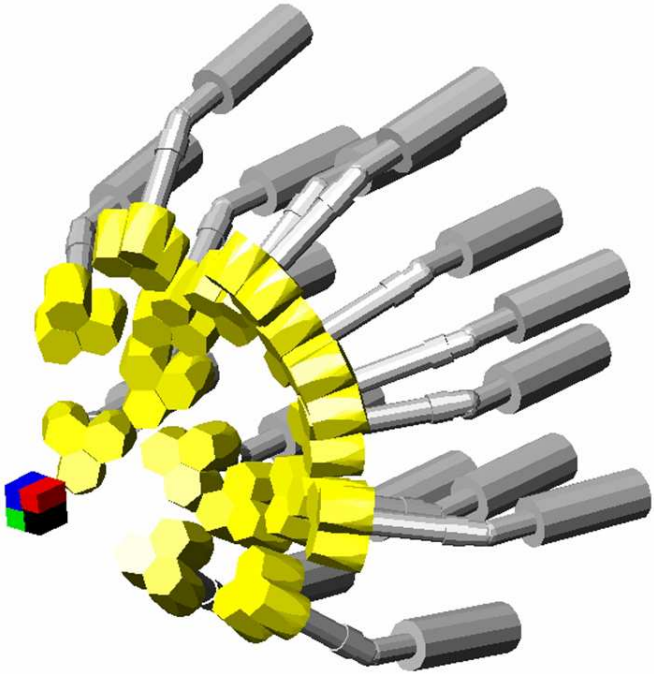}}
  \caption{Hypernuclear physics set-up showing (i) the secondary
    target, and (ii) the cluster detectors, each consisting of 3
    hexagonal HPGe crystals packed in a common cryostat and an
    electromechanical cooler.}
  \label{fig:PANDA}
\end{figure}

The slowing down the $\Xi^-$ proceeds (i) through a sequence of
nuclear elastic scattering events inside the residual nucleus in which
the annihilation has occurred and (ii) by energy loss during the
passage through an active absorber. If decelerated to rest before
decaying, the particle can be captured inside a nucleus, eventually
releasing two $\Lambda$ hyperons and forming a double hypernuclei. In
Fig.~\ref{fig:target} the geometry of the secondary target is shown,
where the radial dimension was determined by the short life-time of
the $\Xi^-$ particle while the angular coverage is limited by the kaon
detection in forward direction, making use of the tracking probability
of the general purpose $\overline{\mbox{P}}$ANDA set-up. The
integration of the hypernuclear physics set-up into
$\overline{\mbox{P}}$ANDA is shown in Fig.~\ref{fig:PANDA}. The
absorber will consist of a sandwich structure of silicon layers and a
combination of target materials. The choice of the target is crucial
for the magnitude of the cross section. Also the number of excited
states of the core should be small and the states should be well
separated.

The level scheme of single and double hypernuclei will be explored by
$\gamma$-ray detection. Only recently, high resolution spectroscopy of
hypernuclei with germanium arrays was established. Spectroscopic
information on double hypernuclei can only be obtained via their
sequential decays. To maximise the detection efficiency the high
purity germanium (HPGe) detectors have to be located near the target,
and thus have to be operated in a strong magnetic field. In
combination with the high luminosity of the HESR machine at FAIR and
with the proposed solid-state micro-tracker, high resolution
$\gamma$-ray spectroscopy of double hypernuclei will become possible
for the first time.

\begin{figure}
  \resizebox{\columnwidth}{!}{
    \includegraphics{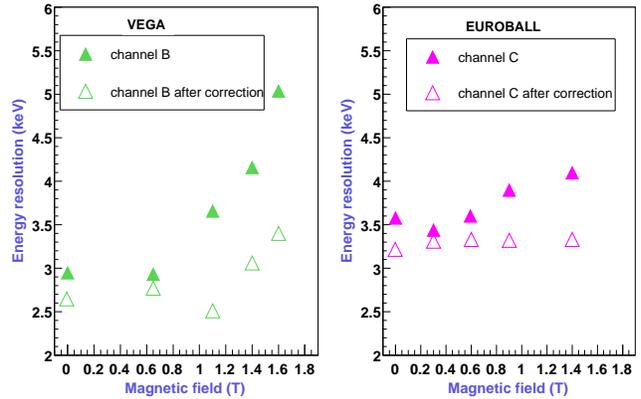}}
  \caption{Measured energy resolution (FWHM) for the 1.333\,MeV
    $\gamma$-line of Euroball cluster and VEGA detectors in a magnetic
    field with and without correction~\cite{SanchezSem}. Statistical
    errors for the values are smaller than the symbol size.}
  \label{fig:VegaEuroball-FWHM}
\end{figure}

The multiple steps in the production and detection chain of double
hypernuclei have been discussed in Ref.~\cite{Pochodzalla2004}. One of
the unknown factors in the realization of the experiment is the
performance of HPGe detectors in strong magnetic fields. To verify
that HPGe detectors can be safely and efficiently operated in such an
environment two different kind of detectors have been tested in the
field provided by the ALADiN magnet at GSI: the Euroball cluster
detector~\cite{Eberth1996} and the VEGA detector~\cite{Gerl1994}. A
small degradation of the energy resolution was found, and a change in
the rise time distribution of the preamplifier pulses was
observed~\cite{Sanchez2007}.  Fig.~\ref{fig:VegaEuroball-FWHM} shows
the energy resolution (FWHM) at $1.333\,$MeV of one channel of each
detector as a function of the magnetic field. A correlation between
rise time and pulse height is used to correct the measured energy,
recovering the initial energy resolution almost completely. This
promising result offers the opportunity not only for hypernuclear
$\gamma$-spectroscopy with $\overline{\mbox{P}}$ANDA, but also for a
possible FINUDA spectrometer up-grade at
DA$\Phi$NE~\cite{Feliciello2007}.

\vspace{1.1cm}

\end{document}